\documentclass[journal]{IEEEtran}

\usepackage{algorithm}
\usepackage{caption}
\usepackage{algpseudocode}
\usepackage{amssymb}

\usepackage{graphicx}
\usepackage{epstopdf}
\usepackage{subfigure}
\usepackage{multirow}
\usepackage{amsthm}
\usepackage{amsmath}
\usepackage{setspace}
\newtheorem{defin}{Definition}

\usepackage{tablefootnote}

\usepackage[numbers,sort&compress]{natbib}

\begin{document}

\title{FASTCloud: A novel framework of assessment and selection for trustworthy cloud service}

\author{Xiang Li~
	\thanks{Xiang Li is with Informatization Construction and Management Office, Sichuan University, Chengdu, Sichuan Province, 610065, China (e-mails: xiangli\_icmo@scu.edu.cn).
}}

\maketitle

\begin{abstract} 

By virtue of technical and cost-effective advantage, cloud computing has increasingly attracted numerous potential cloud consumers plan to adopt cloud service and migrate the traditional IT system to the cloud platform. However, trust has become one of the most challenging issues that prevent them from adopting cloud service, especially in trustworthy cloud service selection. Besides, due to the diversity and dynamic of quality of service (QoS) in the real cloud environment, the existing trust assessment methods based on the single and constant value of QoS attribute as well as the subjective preference based weighting approach are insufficient in efficiency and accuracy. Thus, there is an urgent need to provide a novel and practicable solution for potential cloud consumers to assess and select a trustworthy cloud service among a wide range of functionally-equivalent cloud services. To this end, a novel assessment and selection framework for trustworthy cloud service, FASTCloud, is proposed. This framework can facilitate potential cloud consumers to assess and select a trustworthy cloud service based on their actual  requirements for QoS attributes. In order to improve the accuracy and efficient of cloud service trust assessment, a QoS based trust assessment model is proposed and adopted to the framework. This model presents a trust level evaluation method based on the interval multiple attributes to adaptively evaluate the trust level of various cloud services. Furthermore, an objective weight assignment method based on the deviation maximization is proposed to improve the accuracy of this model. To evaluate the effectiveness and efficiency of FASTCloud, a case study with a real-world dataset and a simulation experiment for performance analysis and comparison are conducted, respectively. Experimental results show that the proposed framework can effectively evaluate the trust level of various cloud services with multiple QoS attributes while the evaluation method in the proposed model efficiently  outperforming other trust assessment methods to a certain extent.
\end{abstract}

\begin{IEEEkeywords}
	trustworthy internet of vehicles, trust model, trust assessment, quality of service
	
\end{IEEEkeywords}

\section{Introduction}
Cloud computing has become a new utilization paradigm of IT resources that provides web-based on-demand services to customers over the internet. Depending on the diverse business requirements of different IT customers, cloud computing offers a variety of service models including infrastructure-as-a-service, platform-as-a-service, software-as-a-service and etc. \cite{}. Compared with the traditional way of investing huge amounts of capital to purchase IT infrastructure, the economic benefits that cloud computing can bring to an enterprise by virtue of its technological advantages are obvious. Moreover, cloud computing also provides the basic platform for the rapid development of other emerging technologies, such as big data and 5G \cite{}, mobile edge computing \cite{} and IoT \cite{li2019enhancing}. In addition, cloud computing can free enterprises from the low-level task of building IT infrastructure so that they can focus more on the high-level task of business innovation to create value for their customers \cite{buyya2018manifesto}. Therefore, more and more organizations and individuals have been experimenting with building business applications on the cloud and making it more agile by adopting flexible and resilient cloud services.

However, it is not easy for the potential cloud customers (PCC), such as enterprises, organizations and individuals that plan to adopt cloud service, to take full advantage of cloud computing \cite{Alabool2018Cloud}. Enterprises will face many challenges in migrating applications, workflow and business from traditional IT systems to the cloud platform. These challenges are often related to the specific requirements and characteristics of the existing business of customers, which depend heavily on the quality of service (QoS) of the cloud service provisioned by cloud service provider (CSP) \cite{2019Trust}. Moreover, with the increasing demands of customers, a large number of cloud services with similar functions and features provided by various CSPs  have emerged in the cloud business market \cite{hayyolalam2018systematic}. Different cloud services can satisfy the multiple QoS requirements for different cloud service customers (CSC). Therefore, it has truly brought about a tough challenge for PCCs to select a trustworthy CSP out of a large pool of candidate CSPs with similar offerings \cite{2016Managing}. That is, how to accurately and objectively assess the trust level of cloud services provided by different CSPs has become one of the most challenging issues for PCCs. 

To address the trust issues in cloud service, various researches on QoS-based assessment and selection for trustworthy cloud service have attracted considerable interest. These studies focus on evaluating the trust level of various cloud services provided by different CSPs by utilizing the information or data related to the QoS attributes of cloud services. The trust level is denoted as a quantitative value, which is usually used to represent the comprehensive performance of CPS in providing cloud service with multifaceted capabilities. However, the availability andaccuracy of information or data regarding multiple QoS attributes of cloud service and the applicability and efficiency of trust evaluation methods are still urgent issues to be solved in the existing researches. 

To this end, we propose a novel assessment and selection framework for trustworthy cloud service, FASTCloud, which enhances availability and accuracy of the obtained information regarding QoS attributes of a cloud service. Furthermore, a QoS based trust assessment model is proposed to improve the applicability and efficiency of trust assessment method. The main purpose of FASTCloud is to facilitate PCCs to select a trustworthy cloud service based on their actual requirements for the QoS attributes of cloud service. Following are the prime contributions of the present research work.

\begin{itemize}
	\item A novel assessment and selection framework for trustworthy cloud services based on diverse and dynamic QoS attributes, FASTCloud, is proposed. The FASTCloud can collects available and valid data related to QoS attributes of cloud services. The data consists of the constant agreed values and dynamic monitoring values regarding these QoS attributes submitted by CSPs and CSCs respectively.
	
	\item For the convenience of PCCs to select a trustworthy cloud service, a selection component is designed in FASTCloud to accept assessment requests initiated by PCCs. This component takes the requirements of the PCC for QoS attributes as metrics and takes cloud services provided by candidate CSPs matched against these metrics as objects to be evaluated. The component utilizes the collected information about the QoS attributes to evaluate the trust level of cloud services and return the results to PCCs.
	
	\item To accurately and efficiently evaluate the trust level of cloud services, a QoS based trust assessment model is proposed and implemented by the component. This model presents a trust level evaluation method based on the QoS attribute with interval value to determine the trust level of cloud services provided by candidate CSPs. In order to objectively determine weights to different QoS attributes, a weight assignment method based on the deviation maximization is adopted in the model. 
	
	\item The experiments are conducted in the form of case study and simulation to validate the effectiveness and efficiency of FASTCloud. The experimental result shows that the proposed framework can effectively facilitate PCCs to achieve the purpose of assessment and selection for trustworthy cloud service. The performance of trust assessment model is analyzed and compared in terms of time complexity and simulation experiment to demonstrate its advantages.
\end{itemize}

The rest of the paper is organized as follows. Section 2 discusses the related work. Section 3 introduces the proposed framework. Section 4 details the proposed cloud service trust assessment model and elaborates on the presented trust level assessment method. Section 5 presents the case study. Section 6 presents the experiment and result analysis. Section 7 presents the conclusions of this paper and outlines directions for future work. We have summarized the definitions of the acronyms that will be frequently used in this paper for ease of reference, as shown in Table 1.

\begin{table}\label{Table 1}
	\label{Acronyms}
	\centering
	\caption{Summary of key acronyms}
	\resizebox{\linewidth}{!}{
	\begin{tabular}{lll} 
		\hline
		Acronym &   Definition  \\
		\hline
		CSP      & Cloud Service Provider                 \\
		CSC     & Cloud Service Customer               \\
		PCC     & Potential Cloud Customer            \\ 
		QoS      & Quality of Service                        \\
		SLA     & Service Level Agreement            \\
		SLO     &  Service Level Objective           \\
		TCSC   & Trustworthy Cloud service Selection Component \\
		AMV    & Actual Monitoring Value \\
		TAM    & Trust Assessment Model \\
		\hline
	\end{tabular}}
\end{table}

\section{Related Work}
In recent years, the research on assessment and selection of trustworthy cloud service has attracted considerable interest of many researchers. A variety of trust assessment methods and trust models have been proposed by taking QoS attributes as metrics. Kumar et al. \cite{kumar2021ccs} proposed a novel framework, Optimal Service Selection and Ranking of Cloud Computing Services (CCS-OSSR), which allowed PCCs to compare available service choices based on QoS. The CCS-OSSR utilized the best worst method to rank and prioritize the QoS criteria, and employed TOPSIS approach to obtain the final rank of cloud services. Furthermore, in \cite{kumar2021computational}, the authors the fuzzy {analytic hierarchy process (AHP) method to define the architecture of overall cloud service selection process and calculate the weights of QoS criteria. These calculated criteria weight are utilized with TOPSIS method to evaluate the final rank of cloud service based on their	overall performance.} Sun et al. \cite{SUN2019749} proposed a cloud service selection with criteria interactions framework (CSSCI) for cloud service selection. This framework applies a fuzzy measure and choquet integral to measure and aggregate non-linear relations between criteria, such as latency, response time and availability. Jatoth et al. \cite{Jatoth2018SELCLOUD} proposed a methodology to addresses a hybrid multi-criteria decision-making model involving the selection of cloud services among the available alternatives. This methodology assigns various ranks to cloud services based on the quantified QoS parameters using a novel extended gray technique for order preference by similarity to an ideal solution (TOPSIS) integrated with AHP. In \cite{2017Design}, three multiple criteria decision making (MCDM)-based multi-dimensional trust assessment schemes have been presented, which assess trust level of CSPs by monitoring compliance provided by CSPs against the set SLAs. These schemes adopt three MCDM methods: AHP, TOPSIS and preference ranking organization methods for enrichment evaluations (PROMETHEE) respectively that enable PCCs to determine the trust level of a CSP from different perspectives.  In \cite{yang2017framework}, a novel method was proposed, which employed a multi-QoS-aware cloud service selection strategy and the AHP method to help the PCCs to select the appropriate cloud service. To select the best one out of available cloud services, Shetty and  D'Mello \cite{shetty2015quality}  proposed a service selection algorithm based on the QoS requirements of PCC.

In addition, there are many researchers tend to adopt the service measurement index (SMI) defined by the Cloud services measurement initiative consortium as QoS attributes for assessment and selection of cloud services. The SMI is one of the widely accepted metrics for quality measurement of cloud service. Singh and Sidhu \cite{singh2017compliance} proposed a compliance-based multi-dimensional trust assessment system, which enabled PCCs to determine the trust level of a CSP. This system helped PCCs select an optimal CSP from candidate CSPs that satisfy its desired QoS requirements. Somu et al. \cite{Somu2018A} presented a trust-centric approach for identification of suitable and trustworthy CSPs. This approach employs multiple algorithms for the identification of similar service providers, credibility based trust assessment, selection of trustworthy service providers, and optimal service ranking respectively. A trust assessment framework that uses the compliance monitoring mechanism to determine the trust level of CSPs was proposed in \cite{sidhu2017improved}. The compliance values are computed and then processed using a technique known as TOPSIS to obtain trust level of CSPs.  

In \cite{kumar2018novel}, a computational framework for determining the most suitable candidate cloud service by integrating the analytical hierarchical process (AHP) and Technique for order preference by similarity to ideal solution (TOPSIS). Such a framework used AHP to define the architecture for selection process of cloud services and compute the criteria weights using pairwise comparison. Then, TOPSIS method is used to obtain the final ranking of the cloud service based on overall performance metrics. Tripathi et al. \cite{2017Integration} proposed an improved SMI-based framework for enabling PCCs to select an appropriate CSP according to their QoS requirements. This framework employed the analytic network process (ANP) method for the ranking of cloud services. Yadav and Goraya \cite{Neeraj2018Two} proposed a novel two-way ranking based cloud service mapping framework for PCCs to select a suitable CSP. In this framework, AHP has been used to assess the ranking score of both the CSPs and PCCs by considering the QoS attributes value offered by them as well as desired by their counterpart. In \cite{somu2017computational}, a 3-tier cloud service selection architecture with hypergraph based computational model (HGCM) and minimum distance-helly property (MDHP) ranking algorithm in the service ranking layer was proposed to measure and quantify the SMI attributes thereby facilitating the PCCs to rank the cloud services. HGCM enables the CSPs to analyze themselves by comparing with other CSPs and to enhance the level of satisfaction experienced by the PCCs.  Moreover, some researchers attempt to assess and select a trustworthy cloud service from the perspective of security, but they lacked an effective and feasible method used by PCCs \cite{silva2019calculating,li2020}.

As aforementioned, there are two deficiencies in the existing studies. On the one hand,  the data or information used by existing studies to evaluate the QoS attributes of cloud service trust level is usually obtained in the form of documents, which are either from the technical specifications or SLA statements on cloud service provided by CSP. It is assumed that the CSP will honestly abide by its commitment in the documents to provide cloud service. However, some CSPs may be driven by profits to exaggerate the QoS of their cloud services to attract more PCCs, which makes the trust assessment of cloud services lack objective fairness and transparency. In addition, it is a challenging issue for individual PCC that effectively obtains the information or data related to QoS attributes of cloud services provided by various CSPs.

On the other hand, trust assessment methods proposed by existing studies usually employ the single and constant value of QoS attributes (e.g., the agreed service level objective (SLO) regarding QoS attribute in the SLA contracted by CSP and CSC) to assess the trust level of cloud services provided by CSPs. In fact, even for the same cloud service and the same QoS attribute, different PCCs may have different  SLO requirements. Then, a CSP must be capable of providing cloud service with various SLOs of QoS attributes for its CSCs. Moreover, the QoS attributes of cloud service are treated theoretically and idealistically as invariant without considering the dynamic performance of cloud service in the real cloud environment. Nevertheless, cloud services may be be affected by the dynamic changes of network, workloads and shared virtualization resources (such as jitter or congestion of network, transactions bursting, capacity expansion and reduction) during operation. It will inevitably lead to the continuous fluctuation of QoS attributes of cloud service, which makes the traditional trust assessment method based on the single and constant value of QoS attributes no longer adopted well to the real and dynamic cloud environment. In practice, during the operation of cloud service, the actual value of its QoS attributes is dynamic and uncertain. Therefore, the trust assessment method of cloud services should be designed from the perspective of  the dynamicity and variability of QoS attributes. 

Furthermore, most of the existing trust assessment methods adopt subjective preference based weighting approach to assign weights for QoS attributes.  Such a method not only affects the accuracy of trust assessment due to the lack of objectivity and flexibility, but also does not apply to the PCCs who do not have professional knowledge and experience in the field of cloud evaluation.

To the best of our knowledge, there is still a lack of effective solutions to tackle with the above issues. Contrary to this, a novel assessment and selection framework for trustworthy cloud service and an efficient trust assessment model are proposed to solve these issues.


\section{The Proposed Framework}
This section proposes an assessment and selection framework for trustworthy cloud service (FASTCloud), which is an extension base on our previous works \cite{Li2018SCCAF}.  The FASTCloud collects the SLO and AMV regarding the QoS attributes of cloud services submitted by CSPs and their CSCs respectively, and utilizes the trust assessment model to evaluate trust level of the cloud services accordingly. The QoS attributes of cloud service provided by a CSP to its CSC are determined and agreed in the SLA contracted by both of them. In practice, the SLA is widely used to define a formal contract between a service provider and a service consumer, which defines the quality level of the service expected from the former and the commitment of the latter. In the cloud context, a CSP manage virtualized IT resources (e.g., computing, storage, network, data, etc.) and provide them to its CSCs in compliance with the SLA.

Moreover, the QoS attributes of cloud service are usually specified in SLA, which defines the SLOs they must meet. For instance, the SLO agreed by CSC and CSP on the response time (i.e., a QoS attribute related to network status, which represents the time taken to send a service request and receive a response) of cloud service in SLA is 100 ms. For that reason, SLA is extensively adopted to ensure that the QoS of cloud service delivered by a CSP conforms to the expectation of a CSC. For ease of illustration, the trust assessment related terms used throughout this paper are defined based on industry standard~\cite{ISO190861}, technical specification~\cite{2008quality} and literature~\cite{huang2013trust}, as shown in Table 2.

\begin{table}
	\label{Glossary}
	\centering
	\caption{Glossary of important trust assessment related terms}		\renewcommand\tabcolsep{1.8 pt}
	\resizebox{\linewidth}{!}{
	\begin{tabular}{ll} 
		\hline
		Term &  Definition  \\
		\hline
		SLA	  & \begin{tabular}[c]{p{0.9\columnwidth}} 
			It is a legally documented agreement between the CSP and CSC used to govern the QoS that the covered service is expected to be provisioned, which includes cloud SLOs for the covered cloud service. It describes the relationship and roles of both parties, and defines the obligations and guarantees of QoS borne by the CSP in case of violations. 
		\end{tabular} \\
		SLO  & \begin{tabular}[c]{p{0.9\columnwidth}} 
			It is a quantitative commitment made by a CSP for a specific QoS attribute of its cloud service, where the value follows the interval scale\tablefootnote{continuous scale or discrete scale with equal sized scale values and an arbitrary zero.} or ratio scale\tablefootnote{continuous scale with equal sized scale values and an absolute or natural zero point.}. It aims at specifying quantifiable QoS attributes for the covered service under cloud context based on mutual understandings and expectations. 
		\end{tabular} \\
		QoS   &\begin{tabular}[c]{p{0.9\columnwidth}} 
			It represents the totality of measurable attributes of a cloud service that bear on its ability to satisfy the stated requirements of a CSC, which aims to implement the concept of measured cloud service. The QoS are considered to be related to the non-functional quality attributes of a cloud service.
		\end{tabular} \\
		Trust     & \begin{tabular}[c]{p{0.9\columnwidth}} 
			It represents a subjective notation of the relationship between CSC and CSP in cloud context. Such a relationship comprises that the CSC expects a specific behavior from the CSP (such as providing cloud service in compliance with SLA), and believes that the expected behavior occurs based on the evidence of the CSPs' competence (such as ensuring the high-level QoS and sufficient resource provision for cloud service), and will to take risk for that belief.     
		\end{tabular} \\
		Trust Level     & \begin{tabular}[c]{p{0.9\columnwidth}} It is a quantitative value of  the "Trust", which represents the comprehensive degree of compliance of a CSP to the promised SLO regarding QoS attributes of cloud service provided to its CSC as per SLA.        
		\end{tabular} \\
		\hline
	\end{tabular}}    
\end{table}

The FASTCloud mainly consists of three entities and a trustworthy cloud service selection component (TCSC), as shown in Figure 1. The main entities in FASTCloud are CSPs, CSCs and PCCs. TCSC is responsible for evaluating trust level of cloud services based on the collected QoS attributes information by employing the trust assessment model, and returning the trust assessment results to PCCs. The specific roles and responsibilities of entities are as follows.

\begin{itemize}
	\item \textbf{CSP} signs an SLA with its CSC on the specific SLO of QoS attributes of the cloud service. CSP operates and maintains cloud services to its CSC in accordance with the SLA. In addition, CSP provides TCSC with SLO of QoS attributes of its cloud service according to the SLA. 
	
	\item \textbf{CSC} signs an SLA with its CSP on the specific SLO of QoS attributes of the cloud service according to QoS requirements of its actual business. Furthermore, CSC monitors the QoS attributes according to the SLA and provides actual monitoring value (AMV) to TCSC during the cloud service runtime.
	
	\item \textbf{PCC} is a requester for a cloud service assessment, that is, a customer planning to purchase and use a cloud service. PCC initiates an assessment request to TCSC based on its QoS requirements and receives assessment results from TCSC (i.e., candidate CSPs), and selects the most trustworthy one among them. 
\end{itemize}

\begin{figure*}\label{Figure 1}
	\centering
	\includegraphics[width=15 cm]{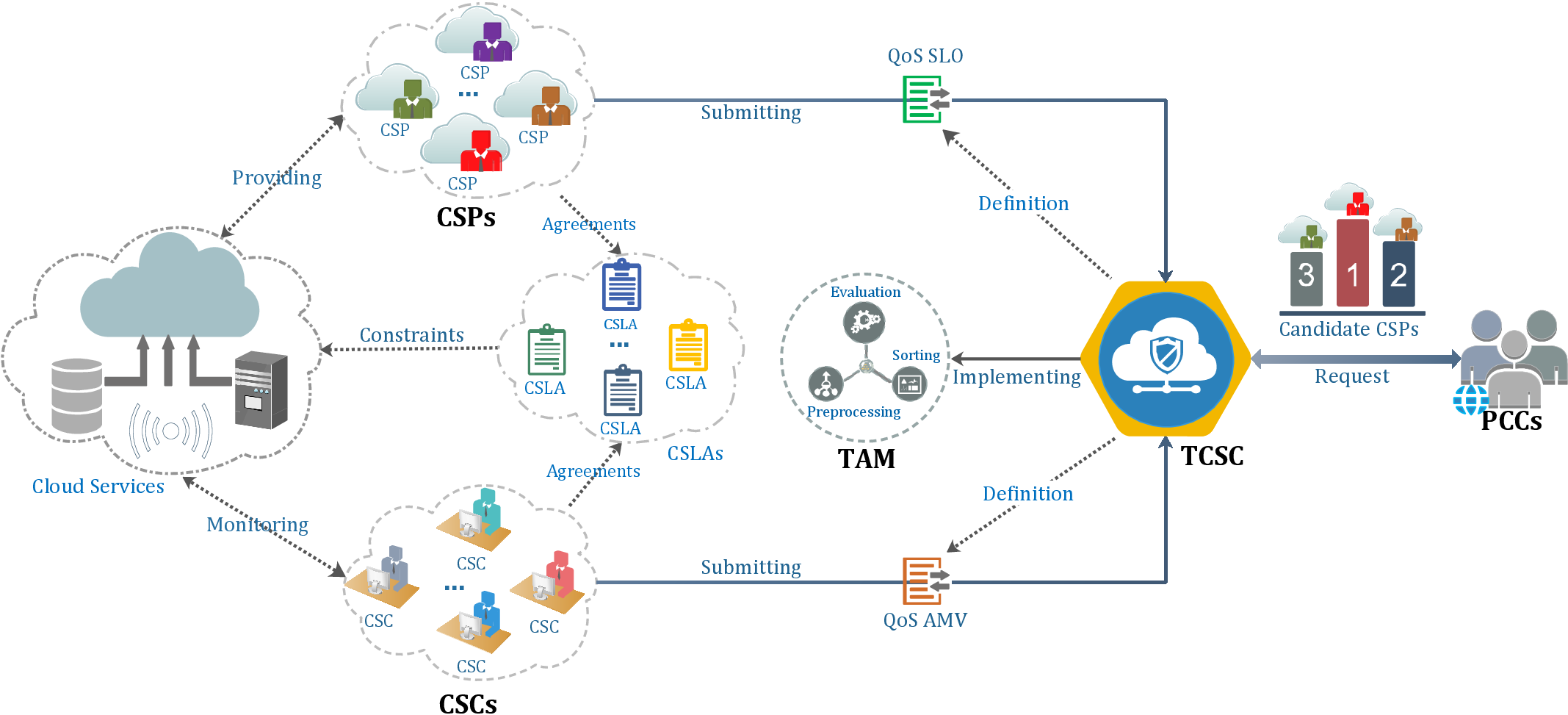}
	\caption{The proposed framework: FASTCloud.}
\end{figure*}

The main role of TCSC is to collect QoS attributes information provided by CSP and CSC (i.e., SLO and AMV) and to assess cloud services. According to the assessment request of PCC and the collected QoS attributes information, TCSC utilizes the trust assessment model (will be detailed later) to assess the trust level of cloud services. Then, TCSC offers the trust assessment results to PCC so that it can select a trustworthy cloud service. The functions and activities of TCSC will be described as follows.

\begin{enumerate}
	\item CSPs submit SLOs of QoS attributes to TCSC. According to the SLA signed with different CSCs, a CSP provides TCSC with SLO of QoS attributes of the cloud service in the form of a uniform specification (e.g., a standard template defined by TCSC). The time with which the CSP provides SLOs of QoS attributes is determined by the frequency of changes in the content of the SLA. For instance, each time a CSP signs a SLA with a new CSC or makes a change (e.g., addition, deletion, or modification) to an existing SLA, it shall provide TCSC with the latest SLOs of QoS attributes of the cloud service.
	
	\item CSCs submit AMVs of QoS attributes to TCSC. In accordance with the SLA signed with a CSP, a CSC continuously monitors the QoS attributes. CSC then provides TCSC with AMV of QoS attributes of the cloud service in the form of a uniform specification (e.g., a standard template defined by TCSC). Since the monitoring tools or services used by different CSCs are various, the time with which the CSC provides AMVs of QoS attributes is determined by itself.  In order to improve the feasibility of AMVs collection, the CSC shall satisfy the principle of minimum submission frequency stipulated by TCSC (e.g., at least once a day). 
	
	\item PCCs initiate a trust assessment request to TCSC for a trustworthy cloud service. When a PCC initiates a trust assessment request along with QoS requirements to TCSC, then TCSC would match the QoS requirements of PCC with the QoS attributes information provided by CSPs and find a list of candidate cloud services which satisfy the QoS requirements of PCC. The TCSC uses the trust assessment model to assess the trust level of the candidate cloud services, and offers a ranked list of trustworthy CSPs to the PCC. 
\end{enumerate}

In fact, compared with the traditional service-oriented computing environment, QoS attributes information in cloud environment is easier to obtain \cite{2013QoS}.  Since most CSPs are generally able to provide monitoring tools/services with free or paid for CSCs to monitor QoS status of their cloud services (e.g., the AWS CloudWatch \cite{CloudWatch}, Microsoft Azure Monitor \cite{AzureMonitor}, Huawei Cloud Eye \cite{CloudMonitorService} and etc.), CSCs can easily acquire the actual value of QoS attributes. Therefore, we assumes that the monitoring tools/services provided by CSPs are trustworthy, so that the actual values of QoS attributes monitored and acquired by CSCs are true. Thus, the AMVs of QoS attributes submitted by CSCs are reliable. In addition, there are also many mature applications and tools (e.g., Web-based interactive online information collection, questionnaire and etc.) that can facilitate TCSC to collect the QoS attributes information provided by CSPs and CSCs.

Therefore, the technical implementation details related to the specific monitoring and collection of QoS attributes information beyond the research scope of this paper, which would not be  discussed further. The rest focuses on the trust assessment model of TCSC, which will be elaborated.

\section{The Trust Assessment Model} 
In this section, a trust assessment model (TAM) is proposed. Such a model employs the data of QoS attributes collected by the TCSC to assess the trust level of cloud services provided by various CSPs. The quantitative assessment results of TAM can facilitate PCCs select a trustworthy cloud service based on their QoS requirements. For convenience, we summarize the major notations in Table 3.

\begin{table}\label{Table 3}
	\label{Notations}
	\centering
	\caption{Summary of major notations}
	\resizebox{\linewidth}{!}{
	\begin{tabular}{ll} 
		\hline
		Notation &   Explanation  \\
		\hline
		$C$      & set of CSPs                 \\
		$U$     &  set of CSCs               \\
		$U_i$  & the CSCs set of $i$th CSP \\
		$A$     & set of QoS attributes of cloud services      \\ 
		$S$      & SLO set of $A$                       \\
		$Q$     & AMV set of $A$                        \\
		$c_i$   & the $i$th CSP in $C$          \\
		$u_ij$  & the $j$th CSC in $U_i$           \\
		$a_k$   & the $k$th QoS attribute in $A$ \\
		$s_{ij}(a_k)$   & the SLO of $a_k$ agreed by $u_ij$ and $c_i$ \\
		$q_{ij}(a_k)$   & the AMV of $a_k$ provided by $u_ij$ \\
		$\lambda_{ik}$  & the consistency rate of $a_k$ of cloud service provided by $c_i$\\
		$\boldsymbol{\omega}$   & weight vector of $A$         \\
		$\omega_k$  & the weight of $a_k$ in $\boldsymbol{\omega}$\\
		\hline
	\end{tabular}}
\end{table}

\subsection{Model Definition}
Let $ C = \left\lbrace c_i | 1\leq i\leq I\right\rbrace $ denote the set of CSPs that provide cloud service to CSCs. Let $ U = \left\lbrace u_{ij}\in U_i | 1\leq i\leq I, 1\leq j\leq J\right\rbrace $ denote the set of CSCs that employ cloud service, where $ U_i $ represents the set of CSCs which employ the cloud service of the $i$th CSP, $u_{ij}$ represents the $j$th CSC of the $i$th CSP. Let $ A=\left\lbrace a_k | 1\leq k \leq K\right\rbrace $ denote the set of QoS attributes of cloud services with the same function and service mode. Let $ S=\left\lbrace s_{ij}\left( a_k\right)\right\rbrace $ denote the SLO set of the QoS attributes, where $s_{ij}(a_k)$ represents the SLO of $a_k$ agreed by the $i$th CSP and its $j$th CSC. Let $ Q=\left\lbrace q_{ij}\left( a_k \right) | 1\leq i\leq I, 1\leq j\leq J,1\leq k\leq K \right\rbrace $ denote the AMV set of QoS attributes, where $q_{ij}(a_k)$ represents the AMV of $a_k$ provided by the $ j $th CSC of the $ i $th CSP.

\subsection{Normalization Processing of QoS Attributes}
For convenience of elaboration, we take the cloud service of a CSP as an example to describe the normalization of QoS information in detail. We assume that for a given CSP (denoted as $ c_i $, $ c_i \in C $), it provide the cloud service with the QoS attributes (denoted as $A$)  to its CSCs (denote as $U_i $). The CSP $ c_i $ and each of its CSCs (denote as $ u_{ij} $, $ u_{ij}\in U_i $) respectively submits the SLO (denoted as $ s_{ij}(a_k), s_{ij}(a_k)\in S$) and the AMV (denoted as $q_{ij}(a_k), q_{ij}(a_k) \in Q$) regarding each of the QoS attribute $a_k, a_k \in A$ to TCSC. Thus, the TCSC can obtain the AMV set (denoted as $ \boldsymbol{S_i(A)}$) and SLO set (denoted as $ \boldsymbol{Q_i(A)} $) regarding the QoS attributes $A$ of the cloud service provided by $c_i$, which can be represented as the following matrix.

\begin{equation}\label{SLOs Matrix of CSP}
\boldsymbol{S_i(A)}=\begin{bmatrix}
s_{i1}(a_1) & s_{i1}(a_2) & \cdots & s_{i1}(a_k)\\
s_{i2}(a_1) & s_{i2}(a_2) & \cdots & s_{i2}(a_k1)\\
\vdots & \vdots & \ddots & \vdots\\
s_{ij}(a_1) & s_{ij}(a_2) & \cdots & s_{ij}(a_k)\\
\end{bmatrix}
\end{equation}
where, $ s_{ij}(a_k) (s_{ij}(a_k)\in S)$ denotes the SLO of  the QoS attribute $a_k$ agreed by $c_i$ and its $j$th CSC.

\begin{equation}\label{AMVs Matrix of CSP}
\boldsymbol{Q_i(A)}=\begin{bmatrix}
q_{i1}(a_1) & q_{i1}(a_2) & \cdots & q_{i1}(a_k)\\
q_{i2}(a_1) & q_{i2}(a_2) & \cdots & q_{i2}(a_k)\\
\vdots & \vdots & \ddots & \vdots\\
q_{ij}(a_1) & q_{ij}(a_2) & \cdots & q_{ij}(a_k)\\
\end{bmatrix}
\end{equation}
where, $ q_{ij}(a_k) $ ($ q_{ij}(a_k)\in Q$) denotes the average AMV of $a_k$ provided by $u_{ij}$.

Since different CSCs have various monitoring frequencies for different QoS attributes, we take the average monitoring value regarding each of the QoS attributes submitted by CSCs as its unique AMV in order to unify measurement benchmark of the QoS attributes. We assume that for a given CSC $u_{ij}$, he submitted $ N $ monitoring values on the QoS attribute $a_k$,  then the average AMV $q_{ij}(a_k)$ of $a_k$ can be obtained by the following equation.

\begin{equation}\label{Mean of AMV}
q_{ij}(a_k)=\dfrac{\sum_{n=1}^{N}q_{ij}(a_k)^{n}}{n}
\end{equation}
where, $q_{ij}(a_k)^{n}$ represents the $n$th monitoring value on $a_k$ submitted by $ u_{ij} $. 

It should be noted that the submission frequency $N$ of monitoring values on the same QoS attribute by various CSCs can be different.

Besides that, in order to accurately and objectively evaluate the trust level of cloud service provided by a CPS, we utilize the AMV submitted by its CSCs to properly calibrate the SLO submitted by the CSP. Its purpose is to alleviate the problem to a certain extent that CSP may be driven by profits to exaggerate SLO regarding the QoS attributes of its cloud services. Consequently, in order to obtain the objective and real SLO regarding the QoS attributes $A$ of cloud service provided by $c_i$, we define the consistency of QoS attributes as follows.

\begin{defin}
	For a given QoS attribute $ a_k $, if its SLO $ s_{ij}(a_k) $ submitted by the CSP $ c_i $ is not less than its AMV $ q_{ij}(a_k) $ submitted by the CSC $ u_{ij} $, 
	then it is considered that $a_k$ of the cloud service provided by $ c_i $ complies with consistency. 
\end{defin}

Moreover, the QoS attributes of cloud service can be divided into two types according to their features: benefit and cost. The benefit QoS attribute refers to that the higher the value of attribute is, the higher its performance or capability is (e.g., throughput and availability). The cost QoS attribute refers to that the higher the value of attribute is, the lower its performance or capability is (e.g., latency, response time). Therefore, for a benefit QoS attribute $ a_k $ of the cloud service provided by CSP $ c_i $, it complies with the condition of consistency is: $ q_{ij}(a_k) \geq s_{ij}(a_k)$. While for a cost QoS attribute $ a_k $ of the cloud service provided by $ c_i $,   it complies with the condition of consistency is: $ q_{ij}(a_k) \leq s_{ij}(a_k)$.

In accordance with the consistency definition of QoS attribute, we can give the definition of its consistency rate.

\begin{defin}
For a given QoS attribute $ a_k $ of cloud service provided by CSP $ c_i $, let $ N_i(a_k) $ represent the number of times that $a_k$ complies with the condition of consistency. Let $ |U_i(a_k) | $ represent the number of AMV regarding $a_k$ submitted by its CSCs $U_i$. Then, the consistency rate on $a_k$ of cloud service provided by $c_i$ can be represented by the ratio of $ N_i(a_k) $ and $ |U_i(a_k) | $, which can be denoted as $\lambda_{ik}$ as follows.
\end{defin}

\begin{equation}\label{Consistency Rate}
	\lambda_{ik}=\dfrac{N_i(a_k)}{|U_i(a_k)|}
\end{equation}

According to equation 4, the consistency rate of QoS attributes $ A $ of cloud service provided by $ c_i $ can be obtained, which are denoted as $ \boldsymbol{\Lambda}=\left[ \lambda_{i1}, \lambda_{i1}, \cdots, \lambda_{ik}\right] $. Since $ N_i(a_k)\leq |U_i(a_k)| $, it can be seen that $ 0\leq \lambda_{ik}\leq 1 $.

For the benefit QoS attribute, $ N_i(a_k) $ can be calculated by the following equation.
\begin{equation}\label{Consistency Number of Benefit QoS}
N_i(a_k)=\left\{\begin{matrix}
&\sum_{j=1}^{|U_i(a_k)|}1 , &q_{ij}(a_k)\geq s_{ij}a(k) \\ \\
&0 , &others 
\end{matrix}\right.
\end{equation}

For the cost QoS attribute, $ N_i(a_k) $ is as follows.

\begin{equation}\label{Consistency Number of Cost QoS}
N_i(a_k)=\left\{\begin{matrix}
&\sum_{j=1}^{|U_i(a_k)|}1 , &q_{ij}(a_k)\leq s_{ij}a(k) \\ \\
&0 , &others 
\end{matrix}\right.
\end{equation}

The minimum and maximum SLO about each of QoS attributes submitted by $ c_i $ can be obtained from $S_i(A) $, which are denoted as $ s_i(a_k)^l $ and $ s_i(a_k)^u $ respectively. The SLO of each QoS attribute of the cloud service provided by $c_i$ can be represented as the interval: $ s_i(\tilde{a}_k) =\left[ s_i(a_k)^l,  s_i(a_k)^u\right] $. It represents the SLO extent about each of QoS attributes claimed by $ c_i $ to its $ U_i $ that its cloud service can comply with. In a real cloud environment, we can intuitively feel the approximate SLO about QoS attributes that a cloud service can truly achieve by observing the AMV about these QoS attributes. Therefore, the objective and real SLO extent about the QoS attributes of cloud service that a CSP is capable of offering to its CSCs, denoted as $ \tilde{b}_{ik} $, can be determined by the consistency rate. It can be obtained by the following equation.

\begin{equation}\label{Actual SLO Interval}
\small 
\tilde{b}_{ik}=\lambda_{ik} \times s_i(\tilde{a}_k)=\left[ \lambda_{ik} s_i(a_k)^l, \lambda_{ik} s_i(a_k)^u \right]=\left[ b_{ik}^l, b_{ik}^u \right]
\end{equation}
where, $b_{ik}^l$ and $b_{ik}^u$ respectively denote the actual minimum and maximum SLO of the QoS attribute $a_k$. 

Therefore, the actual SLO extent (i.e., interval value) about QoS attributes $ A $ of cloud service provided by $ c_i $ can be denoted as $ \boldsymbol{\tilde{b}_i}=\left[ \tilde{b}_{i1}, \tilde{b}_{i2}, \cdots, \tilde{b}_{ik} \right] $.

\subsection{Trust Level Evaluation Method}
Assuming that a PCC issues an trust level assessment request with $ K $ QoS attributes to the TCSC, and $ I $ candidate CSPs satisfying the requirements about QoS attributes would be found. Then, TCSC employs TAM to obtain the SLO extent about $K$ QoS attributes of cloud services provided by the $I$ CSPs, and assesses the trust level of cloud services accordingly. The trust level assessment method comprises five steps, which will be described in details as follows.

\subsubsection{Construct the normalized decision matrix}
According to equations (3-1) - (3-7), the actual SLO interval value $ \boldsymbol{\tilde{b}_i}$ about the $ K $ QoS attributes submitted by the $ I $ CSPs can be obtained. Then, the decision matrix composed of $\boldsymbol{\tilde{b}_i}$ can be denoted as $\boldsymbol{B}=(\tilde{b}_{ik})_{I\times K} $. That is, 

\begin{equation}\label{Decision Matrix}
\small
\boldsymbol{B}=\left[ \tilde{\boldsymbol{b_1}},  \tilde{\boldsymbol{b_2}}, \cdots, \tilde{\boldsymbol{b_i}}, \cdots,  \tilde{\boldsymbol{b_I}} \right]^T=\\
\begin{bmatrix}
\tilde{b}_{11} & \tilde{b}_{12} & \cdots & \tilde{b}_{1k}\\
\tilde{b}_{21} & \tilde{b}_{22} & \cdots & \tilde{b}_{2k}\\
\vdots & \vdots & \ddots & \vdots\\
\tilde{b}_{i1} & 	\tilde{b}_{i2} & \cdots & 	\tilde{b}_{ik}\\
\vdots & \vdots & \ddots & \vdots\\
\tilde{b}_{I1} & 	\tilde{b}_{I2} & \cdots & 	\tilde{b}_{IK}
\end{bmatrix}
\end{equation}

Due to  different QoS attributes may belong to different types (benefit and cost) and have different measurement benchmark, there is a lack of comparability between them. In order to eliminate the impact of these problems on the trust assessment results, the decision matrix $ \boldsymbol{B} $ needs to be normalized. 

The normalized decision matrix $ \boldsymbol{B} $ can be denoted as $ \boldsymbol{R}=(r_{ik})_{I\times K} $.  $ r_{ik} $ is also a interval number, denoted as $ r_{ik}=\left[ r_{ik}^l, r_{ik}^u \right]  $, where $ r_{ik}^l $ and $ r_{ik}^u$ can be represented as follows:

\begin{equation}\label{Normalized Decision Matrix of Benefit Type}
r_{ik}^l=\left\{\begin{matrix}
\left. b_{ik}^l \middle/ \sum_{i=1}^{I}b_{ik}^u \right. ,&k\in E_1\\
\left. (1/b_{ik}^u) \middle/ \sum_{i=1}^{I}(1/b_{ik}^l) \right. ,& k\in E_2 
\end{matrix}\right.
\end{equation}

\begin{equation}\label{Normalized Decision Matrix of Cost Type}
r_{ik}^u=\left\{\begin{matrix}
\left.  b_{ik}^u \middle/  \sum_{i=1}^{I}b_{ik}^l  \right. ,&k\in E_1\\ 
\left.  (1/b_{ik}^l) \middle/ \sum_{i=1}^{I}(1/b_{ik}^u) \right. ,& k\in E_2
\end{matrix}\right.
\end{equation}

\subsubsection{Determine the objective weights of QoS attributes}
As previously mentioned, in the real cloud environment, the QoS attributes of cloud service will fluctuate continuously during its operation. However, most of the existing researches on cloud service trust assessment employs the subjective preference based weight assignment method to determine the weights of different QoS attributes \cite{2016Managing,2019Trust,Alabool2018Cloud,2012Trust}. The weights of QoS attributes obtained by such a method are static constants, which cannot well adapt to the dynamic QoS attributes in the real cloud context. To this end, an objective weight assignment method based on the deviation maximization is adopted to determine the weights of QoS attributes. The rationale of this method is that if the difference of values about a QoS attribute provided by all CSPs is smaller, it indicates that the impact of this QoS attribute on trust assessment is smaller. On the contrary, if a QoS attribute can make the difference of values provided by all CSPs about it significantly different, it indicates that this QoS attribute will play an important role in the trust assessment. In particular, if the values about a QoS attribute provided by all CSPs have no difference, it indicates that this QoS attribute will have no impact on the trust assessment. The specific process of this method are as follows.

Supposing that for the given QoS attributes $A$, let $\boldsymbol{\omega}=\left({\omega_1, \omega_2, \cdots,  \omega_k, \cdots, \omega_K} \right) $ be the weight vector of $ A $, where $ \omega_k \geq 0 $ and conforms to the following constraint. 

\begin{equation}\label{Weight Constraint}
\sum_{k=1}^{K}\omega_k^2=1
\end{equation}

Let $ d\left( r_{ik}, r_{fk} \right)=\left\| r_{ik}- r_{fk}\right\| $ be the separation degree  between $ r_{ik} $ and $ r_{fk} $ in the normalized matrix $ \boldsymbol{R} $, where $ \left\| r_{ik}- r_{fk}\right\| =|r_{ik}^l-r_{fk}^l| + | r_{ik}^u-r_{fk}^u |$. Therefore, for a given QoS attribute $ a_k\left( a_k\in A \right) $, let $ D_{ik}(\boldsymbol{\omega}) $ denote the deviation between $c_i$ and other CSPs regarding the separation degree of $a_k$. It can be represented as follows:

\begin{equation}\label{Deviation of a CSP}
D_{ik}(\boldsymbol{\omega})=\sum_{f=1}^{I}\left\| r_{ik}-r_{fk}\right\| \omega_k=\sum_{f=1}^{I}d(r_{ik}, r_{fk})\omega_k
\end{equation}
where, $1\leq i\leq I$ and $ 1\leq k\leq K $.

In addition, let $D_k(\boldsymbol{\omega}) $ denote the total deviation between each CSP and other CSPs regarding the separation degree of $a_k$, which can be represented as follows:

\begin{equation}\label{Total Deviation of a CSP}
D_{k}(\boldsymbol{\omega})=\sum_{i=1}^{I}D_{ik}(\boldsymbol{\omega})=\sum_{i=1}^{I}\sum_{f=1}^{I}d(r_{ik}, r_{fk})\omega_k
\end{equation}

According to the rationale of the deviation maximization method, the weight vector of QoS attributes $ \boldsymbol{\omega} $ should make the total deviation of all CSPs on all QoS attributes. For this purpose, the objective function is constructed as follows.

\begin{equation}\label{Objective Max Function}
\small
max\left( D(\boldsymbol{\omega}) \right)=\sum_{k=1}^{K}D_k(\boldsymbol{\omega})=\sum_{i=1}^{I}\sum_{k=1}^{K}\sum_{f=1}^{I}d(r_{ik}, r_{fk})\omega_k
\end{equation}

Thus,  the calculation of the weight vector of QoS attributes $ \boldsymbol{\omega} $ is equivalent to solving the optimal solution of equation (14) under the constraints of equation (11). It can be solved by the method presented in literature \cite{Xu2010A}, which is denoted as follows.

\begin{equation}\label{Optimal Solution of Oobjective Max Function }
\omega_k=\frac{\sum_{i=1}^{I}\sum_{f=1}^{I}d(r_{ik}, r_{fk})}{\sqrt{\sum_{k=1}^{K}\left( \sum_{i=1}^{I}\sum_{f=1}^{I}d(r_{ik}, r_{fk})^2 \right)}}
\end{equation}

Since the traditional weight vector generally conforms to the normalization constraint,  $ \omega_k $ need to be normalized. That is,

\begin{equation}\label{Weight Vector Normalized}
\omega_k=\frac{\sum_{i=1}^{I}\sum_{f=1}^{I}d(r_{ik}, r_{fk})} {\sum_{k=1}^{K}\sum_{i=1}^{I}\sum_{f=1}^{I}d(r_{ik}, r_{fk})}
\end{equation}

\subsubsection{Calculate the trust level of CSP}
For a given cloud service with $K$ QoS attributes provided by the CSP $c_i$, let $ z_i( \boldsymbol{\omega}) $ represent the trust level of $c_i$. It can be obtained by aggregating the element $ r_{ik} $ of the normalized decision matrix $ \boldsymbol{R} $ with the weight $ \lambda_{k} $ in the weight vector $ \boldsymbol{\omega} $. That is, 

\begin{equation}\label{Integrated Value of QoS Attributes}
z_i(\boldsymbol{\omega})=\sum_{k=1}^{K}\omega_kr_{ik}
\end{equation}

\subsubsection{Construct the possibility degree matrix}
Since the trust level of CSPs is still an interval value (i.e., $ z_i(\boldsymbol{\omega}) $), it is not easy to rank the cloud services of CSPs directly. Therefore, possibility degree comparison approach is used to rank the $ z_i(\boldsymbol{\omega}) $. According to \cite{2002The}, formal definition of possibility degree is as follows:

\begin{defin}
	If both $ \tilde{a} $ and $ \tilde{b} $ are interval numbers, or one of them is interval number, let them be $ \tilde{a}=[a^l, a^u] $ and $ \tilde{b}=[b^l, b^u] $. Let $ l_a$ and $ l_b $ be denoted as $a^u - a^l $ and $ b^u-b^l $, then the possibility degree of $ \tilde{a} \geq \tilde{b} $ can be represented as follow.
	
	\begin{equation}\label{Possibility Degree Matrix}
	p( \tilde{a} \geq \tilde{b})=\frac{min\left\lbrace l_a+l_b, max(a^u-b^l, 0)\right\rbrace}{l_a+l_b}
	\end{equation} 
\end{defin}

\begin{algorithm}
	\begin{spacing}{0.8}
		\caption{Trust Level Evaluation Algorithm}
		\begin{algorithmic}[1]
			\Require 
			The QoS attributes $ A $ specified by PCC.
			\Ensure 
			The priority ranking of candidate CSPs $ \upsilon $.
			\State Matching the candidate CSPs set $ C $ with QoS attributes $ A $.
			\For {each CSP $ c_i \in C $} 
			\State Extracting the SLO set $ S_i(A) $ submitted by $ c_i $ on $ A $;
			\State Extracting the AMV set $ Q_i(A)  $ submitted by the CSCs of $ c_i $ on $A$;
			\State Calculating the actual SLO interval $ b_i(A) $ according to $ S_i(A) $ and $ Q_i(A) $;
			\EndFor 
			\State Constructing the decision matrix $ B $ with $ b_i(A) $;
			\For {each $ b_i(A) \in B $}
			\State Normalizing $ b_i(A) $ to $ r_i(A) $ according to the type of $ a_k $.		
			\EndFor
			\State Constructing the normalized decision matrix $ R $ with $ r_i(A) $;
			\For {each $ r_i(A) \in R $  and each $a_k \in A $}
			\State Calculating the deviation $ D_{ik}(\omega) $ of $ c_i $ on $ a_k $;
			\EndFor
			\For {each $ c_i \in C $}
			\State Calculating the total deviation $ D_{A}(\omega) $ of $ c_i  $ on $ a_k $;
			\EndFor
			\State  Determing the weight vector $ \omega $ of $A$ by solving the optimal problem that maximizes $ D_{A}(\omega) $;
			\State  Obtaining the trust level $ Z $ of $ C $ by aggregating $ R $ with $ \omega $;
			\For {each $ z_i \in Z $}
			\State Calculating the possibility degree $ p_i $ of $ z_i $;
			\EndFor
			\State  Constructing the possibility degree matrix $ P $ with $ p_i $;
			\State  Calculating the ordering vector $\upsilon$ of $ P $;
			\State \textbf{return} {$\upsilon$}; 
		\end{algorithmic}
	\end{spacing}
\end{algorithm}

For the given CSP $c_i$ and CSP $ c_e $, let $z_i(\boldsymbol{\omega})$ and $ z_e(\boldsymbol{\omega})$ denote the trust level of $c_i$ and $c_e$ respectively. Let $p(z_i(\boldsymbol{\omega}) \geq z_e(\boldsymbol{\omega}) ) $ denote the possibility degree of them, which can be represented as $ p_{ie} \left( 1\leq i, e\leq I \ and \ i \neq e \right) $ for short. Then, the possibility degree matrix that contains the possibility degree of pairwise comparison between all candidate CSPs,  denoted as $ P=(p_{ie})_{I\times I} $, can be constructed according to Definition 3. Therefore, the  problem of ranking candidate CSPs based on their trust level can be transformed into the ordering vector problem of the possible degree matrix, which is described below.

\subsubsection{Rank the cloud services of CSPs}
Let $ \boldsymbol{\upsilon}= (\upsilon_1, \upsilon_2, \cdots, \upsilon_i, \cdots, \upsilon_I) $ be the ordering vector of the possible degree matrix $ \boldsymbol{P} $. According to \cite{2002The}, the equation of ordering vector is as follows:

\begin{equation}\label{Ordering Vector }
\upsilon_i=\frac{1}{I(I-1)}\left( \sum_{e=1}^{I}p_{ie} + \frac{I}{2} - 1 \right)
\end{equation}

According to $ \upsilon_i $, the priority of cloud services provided by the candidate CSPs  that satisfy the QoS requirements of PCC can be obtained by ranking $z_i(\boldsymbol{\omega})$. Then, the PCC can select a trustworthy cloud service from the candidate CSPs based on the ranking results. Algorithm 1 illustrates the trust level assessment process.

\begin{table}\label{Table 4}
	\centering
	\label{Definition of QoS attributes}
	\caption{Definition of QoS attributes\cite{QWSdataset2007}}		\renewcommand\tabcolsep{1.8 pt}
	\resizebox{\linewidth}{!}{
	\begin{tabular}{lllll}
		\hline
		QoS attributes & Abbreviation & Unit  & Type & Definition \\ \hline
		Availability     & av       & \%                 & B     & Number of successful invocations/total invocations \\
		Throughput     & th       & invokes/s       & B     & Total Number of invocations for a given period of time \\
		Successability  & su      & \%                   & B      & Number of response/number of request messages         \\ 
		Reliability        & re      &\%                      & B        & Ratio of the number of error messages to total messages \\
		Latency           & la       & ms                     & C         & Time taken for the server to process a given request  \\
		Response Time & res    & ms                     & C          & Time taken to send a request and receive a response\\ \hline
	\end{tabular}}
\end{table}

\section{Case Study}
We conduct a case study by using an open source dataset to validate the availability of TAM. The purpose of the case study is to illustrate the trust assessment process of FASTCloud framework.  The dataset, named as QWS\cite{QWSdataset2007}, consists of 2,507 pieces of real data produced by hundreds of Web services on the 6 QoS attributes. The definitions of QoS attributes contained in QWS are shown in Table 4, where $ B $ and $ C $ in the type column represent the benefit and cost respectively.

The motivation to use the QWS for case study mainly depends on the following considerations.

\begin{itemize}
	\item The data contained in the QWS comes from real-world Web services, which can reflect the actual QoS attributes of these services in the real environment to a certain extent.
	
	\item The QoS attributes of Web services defined in the QWS are also applicable to cloud services in the real environment.
	
	\item Each QoS attribute of Web services in the QWS contains multiple data, which can be well applied to the evaluation scenario of the proposed framework and meet the interval value requirements of TAM for QoS attributes.
\end{itemize}

\begin{figure*}\label{Figure 2}
	\centering
	\includegraphics[width=15 cm]{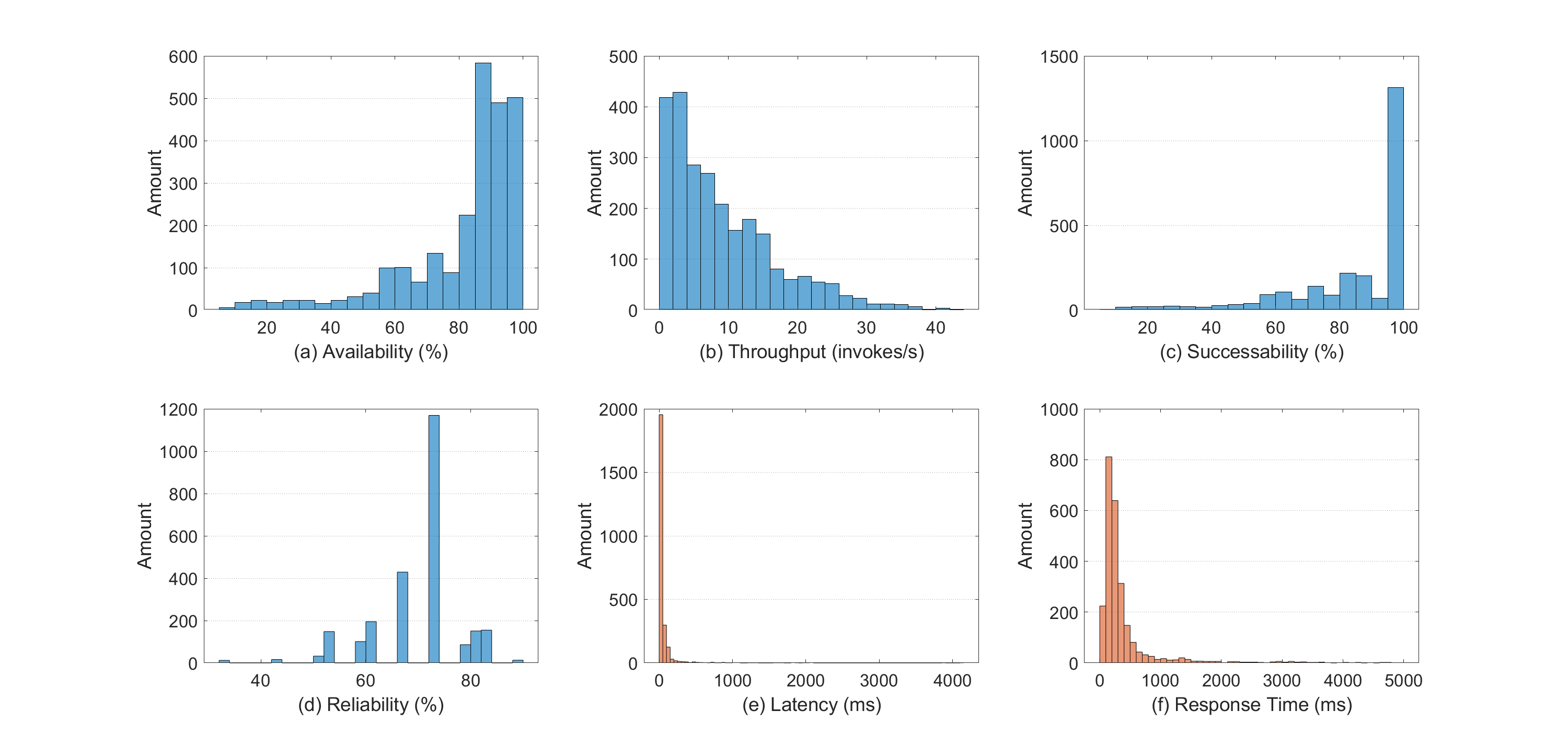}
	\caption{The QoS distribution statistic of QWS.}
\end{figure*}

\subsection{SLO Setup for QoS Attrubites}
For ease of intuitive understanding, we have made statistical analysis on QoS data in the QWS, as shown in Figure 2. As can be seen from Figure 2, the distributions of the QoS attributes values are different, where the distributions of availability, successability, latency and response Time are relatively centralized, while the distributions of throughput and reliability are relatively scattered. In fact, different Web services of QWS contain different amounts of QoS attribute value. For instance, some Web services contain only a value set of QoS attributes (the 6 QoS attributes values represented as a value set), while some Web services contain multiple value sets of QoS attributes. Moreover, there are small number of Web services contain some outliers on the QoS attributes. Therefore, in order to focus on the details of trust level assessment method for cloud service, this experiment simplifies the information processing process of QoS attributes in TAM (as aforementioned in subsection 3.3.2)  and presents the following case study.

\begin{table}\label{Table 5}
	\label{The actual SLO interval value of CSPs on the QoS attributes}
	\renewcommand\tabcolsep{1.8 pt} 
	\centering
	\caption{The SLO interval value of CSPs and PCC on the QoS attributes}
		\resizebox{\linewidth}{!}{
		\begin{tabular}{@{}lllllll@{}} 
			\hline
			\multirow{2}{*}{} & \multicolumn{6}{c}{QoS Attributes} \\ 
			&av &th &su &re &la &res \\
			\hline
			$ CSP_1 $  &[87, 96]  &[6, 23]  &[95, 98]  &[58, 73]  &[8, 33]  &[103, 204]  \\
			$ CSP_2 $  &[62, 97]  &[9, 32]  &[63, 99]  &[56, 83]  &[9, 29]  &[113, 246]  \\ 
			$ CSP_3 $  &[61, 92]  &[4, 26]  &[60, 93]  &[62, 69]  &[7, 27]  &[89, 215]    \\
			$ CSP_4 $  &[71, 78]  &[5, 30]  &[72, 85]  &[59, 67]  &[6, 31]  &[124, 198]  \\
			$ CSP_5 $  &[70, 81]  &[7, 21]  &[69, 82]  &[63, 74]  &[8, 26]  &[92, 193]     \\
			$PCC$       &[50, 100]&[1, 35]  &[50, 100]&[50, 100]&[1, 100]&[50, 300]   \\
			\hline
		\end{tabular}}
\end{table}

We assume that a PCC initiates an assessment request to TCSC and specifies the SLO requirements interval on QoS attributes of cloud services according to the distribution of QoS attributes values in the QWS (i.e., Figure 2), as showed in Table 5. Since it is difficult to obtain the real SLOs on QoS attributes of cloud services provided by CSPs in the real scenario, the SLO interval values of QoS attributes in Table 5 are taken as the agreed SLO of CSPs and CSCs. The QWS dataset is used as the AMV on the QoS attributes of cloud services submitted by the CSCs of these CSPs.  In addition, assume that TCSC matched 5 candidate CSPs satisfy the SLO requirements of the PCC according to its assessment request, denoted as $ CSP_1 $,  $ CSP_2 $, $ CSP_3 $, $ CSP_4 $ and $ CSP_5 $. The maximum and minimum values of these candidate CPSs on each QoS attribute are taken as their actual SLO interval, as shown in Table 5.
 

\subsection{Trust Level Assessment for Cloud Services}
Thus, according to the actual SLO interval value of candidate CSPs on QoS attributes, the trust level of cloud services of each candidate CSPs can be obtained by employing TAM. The specific process are described as follows.

First, the normalized decision matrix of candidate CSPs $ \boldsymbol{R} $ can be constructed by the data of Table 5 and equations (8) - (10), denoted as follows.

\begin{align}
	\tiny
	\setlength{\arraycolsep}{0.5pt}
\boldsymbol{R}=\begin{pmatrix}
&[0.196, 0.274] &[0.0465, 0.742] &[0.208, 0.273]  &[0.159, 0.245]  &[0.0452, 0.725]  &[0.101, 0.407]\\
&[0.14, 0.276]   &[0.0698, 0.968] &[0.138, 0.276]  &[0.153, 0.279]  &[0.0514, 0.644]  &[0.0834, 0.371] \\ 
&[0.137, 0.262] &[0.031, 0.839]   &[0.131, 0.259]  &[0.169, 0.232]  &[0.0552, 0.828]  &[0.0955, 0.417] \\
&[0.16, 0.222]   &[0.0388, 0.936] &[0.158, 0.237]  &[0.161, 0.225]  &[0.0481, 0.966]  &[0.104, 0.338]\\
&[0.158, 0.231] &[0.0543, 0.677] &[0.151, 0.228]  &[0.172, 0.284]  &[0.0574, 0.725]  &[0.106, 0.456]\\
\end{pmatrix}
\nonumber
\end{align}

Second, the weight of each QoS attribute can be calculated by $\boldsymbol{R}$ and equations (11)-(16). The weight vector $\boldsymbol{\omega}$ of QoS attributes can be obtained, denoted as follows.

\begin{equation}\nonumber
\label{Weight Vector of QoS attributes in Case Study}
\boldsymbol{\omega}=\left( 0.0295 \quad 0.118 \quad  0.150 \quad  0.167 \quad  0.247  \quad  0.288 \right).
\end{equation} 

Third, the trust level of each candidate CSP can be calculated by aggregating $ \boldsymbol{R} $ with $\boldsymbol{\omega}$.

\begin{align*}
\label{Trust Level of CSPs in Case Study}
z_1(\boldsymbol{\omega})&=\left[0.109, 0.474 \right], z_2(\boldsymbol{\omega})=\left[0.0953, 0.477 \right],\\
z_3(\boldsymbol{\omega})&=\left[0.0968, 0.525 \right],
z_4(\boldsymbol{\omega})=\left[0.102, 0.526 \right],\\
z_5(\boldsymbol{\omega})&=\left[0.107, 0.473 \right].
\end{align*}

Then, the possibility degree matrix $ \boldsymbol{P} $ of candidate CSPs can be constructed by equation (18) based on their trust level.

\begin{equation}\nonumber
\label{Possibility Degree Matrix of CSPs in Case Study}
\boldsymbol{P}=\begin{pmatrix}
&0.5       &0.508   &0.476  &0.472   &0.502  \\
&0.493   &0.5       &0.469  &0.465   &0.494  \\
&0.524   &0.531   &0.5      &0.496   &0.526  \\
&0.528   &0.535   &0.504  &0.5       &0.53   \\
&0.498   &0.506   &0.474  &0.47     &0.5     \\
\end{pmatrix}
\end{equation}

Finally, the ordering vector of $ \boldsymbol{P} $ can be calculated by equation (19), represented as follows.

\begin{equation}\nonumber
\label{The Ordering Vector of CSPs in Case Study}
\boldsymbol{\upsilon}=\left(0.198 \quad  0.196 \quad  0.204 \quad  0.205 \quad  0.197 \right).
\end{equation}

Therefore, the priority ranking of candidate CSPs can be obtained by sorting the components of $ \boldsymbol{\upsilon} $. That is,

\begin{equation}\nonumber
CSP_4 \mathop{\succ}\limits_{0.504} CSP_3 \mathop{\succ}\limits_{0.524} CSP_1 \mathop{\succ}\limits_{0.502} CSP_5 \mathop{\succ}\limits_{0.506} CSP_2.
\end{equation}
It can seen that $CSP_4$ is the best and $CSP_2$ is the worst.

Similarly, the priority ranking of candidate CSPs can be obtained according to the different types of QoS attributes, as shown in Figure 3. As can be seen from Figure 3, the CSPs priority ranking obtained according to the benefit QoS is different from that obtained based on the cost QoS. For the cost QoS, the priority ranking of candidate CSPs is: $CSP_4 >CSP_3>CSP_5>CSP_1>CSP_2$. For the benefit QoS, the priority ranking of candidate CSPs is: $CSP_2 >CSP_4>CSP_1>CSP_3>CSP_5$. 

\begin{figure}\label{Figure 3}
	\centering
	\includegraphics[width=9 cm]{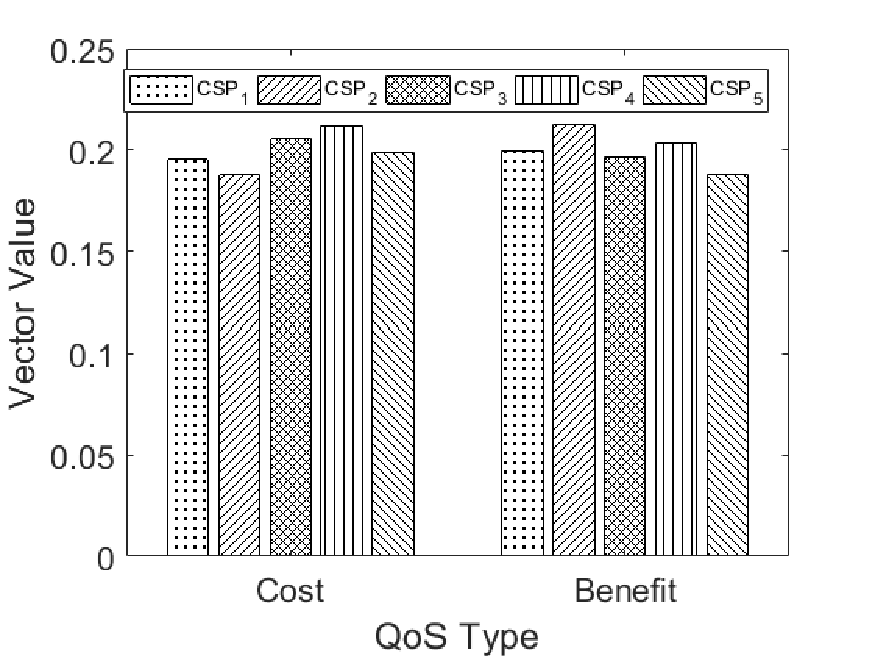}
	\caption{The priority ranking of candidate CSPs based on cost and benefit QoS.}
\end{figure}


\section{Experiment and Result Analysis}
In this section, we carry out complexity analysis and simulation experiments from the perspective of theory and practice to verify the efficiency of TAM. The performance of TAM is analyzed and compared with the traditional and pervasive AHP-based assessment or selection methods for cloud services. First, in order to theoretically verify the efficiency of TAM, we analyze and compare each stage of cloud service assessment process proposed by the competitive methods in the form of time complexity. Then, in order to simulate the efficiency of TAM in the real environment and further validate the accuracy of theoretical analysis, a simulation experiment is carried out to analyze and compare the performance of the competitive methods in the form of time consumption. 

\subsection{Complexity Analysis}
The cloud service assessment process of TAM is divided into five steps, and the time complexity of each step is related to the number of CSPs and QoS attributes to be assessed. Algorithm 1 illustrates the process of TAM. Therefore, it is assumed that the number of CSPs is $ m $ and the number of QoS attributes is $ n $. The time complexity of TAM is analyzed step by step. 

\begin{itemize}
	\item \textbf{Step 1: }In this step, a decision matrix consisting of the actual SLO values of $ m $ CSPs on the $ n $ QoS attributes needs to be constructed and normalized. Since the $ n $ QoS attributes may belong to different types (i.e., benefit and cost), in the worst case, the elements $ r_{ik} $ in the decision matrix $ \boldsymbol{R}_{m\times n} $ need to be normalized by equation (9) and (10). Thus, the time complexity of constructing the normalized decision matrix is: $ O(2mn+2mn)=O(4mn) $.
	
	\item \textbf{Step 2: } The weight vector $ \boldsymbol{\omega} $ of QoS attributes can be determined in this step. Firstly, the total deviation $ D_k(\boldsymbol{\omega}) $ of each CSP from other CSPs on each QoS attribute of normalized decision matrix need to be calculated according to equation (12) and (13). Secondly, the $ D_k(\boldsymbol{\omega}) $ need to be maximized by equation (14). Finally, the weight $ \boldsymbol{\omega} $ of each QoS attribute can be calculated and normalized by equation (15) and (16). Thus, the time complexity of determining the weight vector of QoS attributes is: $ O(2m^2n+2n) $.  
	
	\item \textbf{Step 3: } In this step, the trust level of each CSP $ z_i(\boldsymbol{\omega}) $ can be calculated by aggregating the normalized decision matrix $ \boldsymbol{R} $ with the weight vector $ \boldsymbol{\omega} $. That is,  the integrated value of $ m $ CSPs on $ n $ QoS attributes can be obtained by equation (17). Therefore, the time complexity of calculating the trust level of CSPs is: $ O(mn) $.
	
	\item \textbf{Step 4: }The purpose of this step is to construct the possibility degree matrix of CSPs. The possibility degree $ p_{ie} $ of pairwise comparison between the $ i $th CSP and $ e $th CSP can be calculated by equation (18). The possibility degree matrix  $ \boldsymbol{P}_{I\times I} $ can be constructed on the basis of all $ p_{ie} $. The time complexity of  this step is: $ O(4m^2) $.
	
	\item \textbf{Step 5: }In order to facilitate PCC to select the trustworthy cloud service, the candidate CSPs need to be ranked by the ordering vector of the possibility degree matrix in this step. The ordering vector $ \boldsymbol{\upsilon} $ of $ \boldsymbol{P} $ can be calculated according to equation (19). The priority of $m$ CSPs whose cloud service conforms to the QoS attributes requirements of PCC can be obtained by the ordering vector $ \boldsymbol{\upsilon}$. Therefore, The time complexity of  this step is: $ O(m^2+m) $.
\end{itemize}

In conclusion,  in the worst case, the time complexity of TAM is as follows: $ O(4mn+2m^2n+2n+mn+m^2+m)=O((2n+1)m^2+(5n+1)m+2n) $.

\subsection{Complexity Comparison}
A two-way ranking method based on AHP for cloud service mapping (denote as TRSM) was proposed in literature \cite{Neeraj2018Two}. TRSM divides the QoS requirements of CSCs into multiple criteria layers according to the standard AHP method. Each layer contains different sub-attributes to make it easier to calculate the weight of each attribute and aggregate them accordingly. At the same time, CSPs act as the solution layer where the service quality of their cloud services were assessed and ranked. For a given CSC, the time complexity of TRSM in the worst case are analysed in \cite{Neeraj2018Two}.  That is, $ O\left( \sum_{l=1}^{L}\sum_{i=1}^{N_{n-1}}n_{li}^3+m^3N_l+m\sum_{l=1}^{L}N_l \right) $, where, $ m $ denotes the number of CSPs, $ L $ denotes the number of QoS attributes layers, $ N_L $ denotes the number of Qos attributes contained in each layer, $ n_{li} $ denotes the number of sub-attributes contained in the $ i $th QoS attribute of the $ l $th layer,  which are contained in the $ l-1 $th layer.  In general, the time complexity of TRSM depends on the above four parameters (i.e., $ m $, $ L $, $ N_L $ and $ n_{li} $). However, for the particular cloud services (e.g., with the same functionality and service mode), the hierarchy of their QoS attributes is fixed, then the parameters $ L $ and $ N_L $ are constant. Hence, the time complexity of TRSM is actually determined by the number of CSPs $ m $ and the number of QoS attributes $ n $, which can be represented as $ O(n^3+m^3n+mn) $.

In \cite{singh2017compliance}, a trust evaluation method based on the technique for order preference by similarity to an ideal solution (TOPSIS) and the AHP (denote as AHP-TOPSIS for short) is proposed to evaluate the trust level of CSPs. The QoS attributes were divided into two layers, objective layer and attributes layer. The TOPSIS method acted as the main process to evaluate the trust level of CSPs based on QoS attributes.The AHP method was used to determine the weights of QoS attributes in the main process. For the sake of illustration, the literature let the number of CSPs be $m$ and the number of QoS attributes be $n$, and elaborated the evaluation procedure step by step. According to the step 3 and its sub-steps for the weights assignment of QoS attributes (i.e., AHP was adopted), the time complexity can be roughly calculated and denoted as $O(6n^2+(m+4)n+4m)$. According to the other steps for the trust level evaluation of CSPs based on QoS (i.e., TOPSIS was adopted), the time complexity can be roughly calculated and denoted as $O(2n^2+(m+1)n+m)$. Thus, in the worst case, the time complexity of AHP-TOPSIS is as follows: $O(6n^2+(m+4)n+4m+2n^2+(m+1)n+m) =O(8n^2+(2m+5)n+5m)$.    

Normally,  the largest order of magnitude of the polynomial $O(m, n)$ would be taken as its time complexity. In order to compare the time complexity of TAM, TRSM and AHP-TOPSIS, let $ m $ and $n$ represent the number of CSPs and the number of QoS attributes respectively. We assume that for the given number of QoS attributes, namely $ n $ is constant, the time complexity of TAM, TRSM and AHP-TOPSIS are determined by the number of CSPs $ m $ and respectively denoted as $ O(m^2) $, $ O(m^3) $ and $ O(m) $. Similarly, for the given number of CSPs $ m $, namely $ m $ is constant, the time complexity of TAM, TRSM and AHP-TOPSIS are determined by the number of QoS attributes $ n $ and respectively denoted as $ O(n) $, $ O(n^3) $ and $O(n^2)$. 

Table 6 shows the time complexity comparison of three competitive methods. It can be seen from this table that in the case of a constant number of CSPs, the proposed TAM outperforms TRSM and AHP-TOPSIS. In the case of a constant number of QoS attributes, the time complexity of TAM is between TRSM and AHP-TOPSIS. In order to verify the correctness of the analysis results, simulation experiment are carried out to illustrate the impact of the number of CSPs and the number of QoS attributes on the performance of the three methods.We will further verify the analysis results through the following simulation experiment.

\begin{table}\label{Table 6}
	\label{The time complexity comparison}
	\caption{The time complexity comparison of three competitive methods}
	\resizebox{\linewidth}{!}{
	\begin{tabular}{llll}
		\hline
		\multicolumn{1}{c}{\multirow{3}{*}{Method}} & \multicolumn{3}{c}{Time Complexity:$O(m,n)$\tablefootnote{$m$: the number of CSPs, $n$:the number of QoS attributes.}}                                                     \\ \cline{2-4} 
		\multicolumn{1}{c}{}                        & \multicolumn{1}{c}{\multirow{2}{*}{In the Worst Case}} & \multicolumn{2}{c}{Largest Order of Magnitude} \\ \cline{3-4} 
		\multicolumn{1}{c}{}                        & \multicolumn{1}{c}{}                            & $m$ is a variable      & $n$ is a variable     \\ \hline
		TRSM                                        & $O(n^3+m^3+mn)$                                 & $ O(m^3) $             & $ O(n^3) $            \\
		AHP-TOPSIS                                  & $O(8n^2+(2m+5)n+5m) $                           & $O(m)$                 & $O(n^2)$              \\
		TAM                                         & $O((2n+1)m^2+(5n+1)m+2n)$                       & $O(m^2)$               & $ O(n) $      \\ \hline       
	\end{tabular}}
\end{table}

\subsection{ Simulation Experiment}
The simulation experiment is implemented by using MATLAB R2017b and performed on a DELL desktop computer with configuration as: Intel Core i5 2.7 GHz CPU,8 GB RAM, and Windows 10 operating system. It is assumed that the number of CSPs is $m$ and the number of QoS attributes is $n$. The SLO of QoS attribute is set as a single value and randomly assigned in advance, so as to analyze the impact of the change of $m$ and $n$ on the performance of each evaluation method.The execution time of three competitive methods is the mean of 10 repeated experiment under the same condition, as shown in Table 7 and Table 8.

\begin{table}\label{Table 7}
	\label{The time conmpution comparison 1}
	\caption{The time consumption comparison of three competitive methods under the Condition 1}
	\renewcommand\tabcolsep{1.8 pt} 
	\resizebox{\linewidth}{!}{
	\begin{tabular}{lllllllllll}
		\hline
		\multicolumn{1}{c}{\multirow{2}{*}{Method}} & \multicolumn{10}{c}{\begin{tabular}[l]{@{}l@{}}Execution Time (ms)\tablefootnote{The execution time is the mean of 10 repeated experiment under the Condition 1.}:Condition 1\tablefootnote{The number of CSPs $m=6$ is a constant, the number of QoS attributes $n$ is a variable.}\end{tabular}} \\ \cline{2-11} 
		\multicolumn{1}{l}{}                         & $n=50$         & $n=100$         & $n=150$         & $n=200$          & $n=250$          & $n=300$          & $n=350$          & $n=400$          & $n=450$          & $n=500$         \\ \cline{1-11}
		TRSM                                         & 21.78          & 55.01           & 105.13          & 186.65         & 238.31         & 357.62         & 452.95         & 609.51         & 733.24         & 870.5         \\
		AHP-TOPSIS                                   & 14.31          & 40.22           & 88              & 138.52         & 203.1          & 287.01         & 374.88         & 504.96         & 644.66         & 801.81        \\
		TAM                                          & 11.5           & 21.43           & 32.63           & 43.33          & 59.74          & 84.06          & 115.15         & 127.72         & 135.32         & 159.29        \\ \hline
	\end{tabular}}
		
\end{table}

\begin{table}\label{Table 8}
	\label{The time conmpution comparison 2}
	\caption{The time consumption comparison of three competitive methods under the Condition 2}
	\renewcommand\tabcolsep{1.8 pt}
	\resizebox{\linewidth}{!}{ 
	\begin{tabular}{lllllllllll}
		\hline
		\multicolumn{1}{c}{\multirow{2}{*}{Method}} & \multicolumn{10}{c}{\begin{tabular}[c]{@{}l@{}}Execution Time (ms)\tablefootnote{$^1$The execution time is the mean of 10 repeated experiment under the Condition 2.} :Condition 2\tablefootnote{The number of QoS attributes $n=30$ is a constant, the number of CSPs $m$ is a variable.}\end{tabular}} \\ \cline{2-11} 
		\multicolumn{1}{c}{}                         & $m=6$         & $m=12$         & $m=18$         &  $m=24$            & $m=30$           & $m=36$            & $m=42$            & $m=48$            & $m=54$           & $m=60$           \\ \cline{1-11}
		TRSM     & 8.7          & 31.62           & 41.99       & 72.63     & 100.9    & 187.74     &205.51         & 296.69        &341.12        & 548.64         \\
		AHP-TOPSIS        & 1.12          &2.06           &3.33             & 3.66       & 4.18      & 5.19         & 6.61         & 6.85        & 7.74         & 8.95        \\
		TAM        & 3.58          & 11.57           &25.71           & 45.27          &73.05          & 103.59        &146.24         &197.67        &246.43        &296.11        \\ \hline
	\end{tabular}}
\end{table}

\textbf{Condition 1: } We set the number of CSPs as a constant, namely $m=6$ and increase $n$ from 50 to 500 with a step 50. The experimental result is shown in Figure 4(a).

\textbf{Condition 2: }We set the number of QoS attributes as a constant, $n=30$ and increase $m$ from 6 to 60 with a step 6. The experimental result is shown in Figure 4(b).

\begin{figure*}\label{Figure 4}
	\centering
	\includegraphics[width=15 cm]{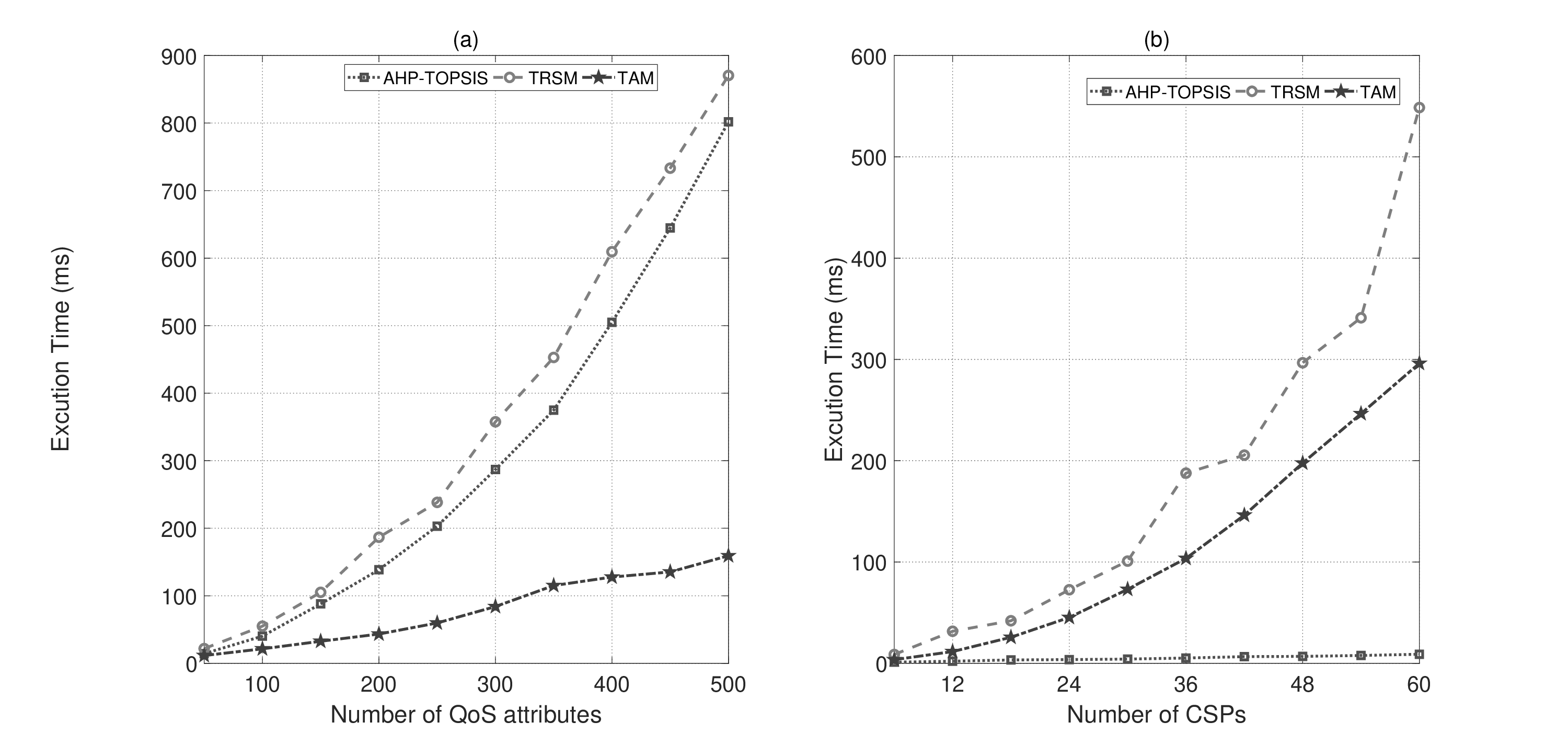}
	\caption{The performance comparison of three competitive methods under two conditions.}
\end{figure*}

Figure 4(a) shows that the execution time of the three competitive methods increases with the number of QoS attributes in the case of the number of CSPs is constant, where TRSM has the largest increase amplitude. Figure 4(b) shows that the execution time of the three competitive methods increase with the number of CSPs in the case of the number of  QoS attributes is constant, where TRSM increase the most, TAM followed, and AHP-TOPSIS increase the least.

The experimental results show that TRSM has the fastest growth rate in execution time under two different conditions, while AHP-TOPSIS and TAM have their advantages and disadvantages respectively. This is because TRSM evaluates the priority of CSPs based on hierarchical QoS attributes. The priority of CSPs is calculated based on each QoS attribute of each layer. The value of different CSPs with respect to each QoS attribute in the upper layer can be obtained by aggregation. Then the above steps are repeated until the priority of CSPs at the highest QoS level is obtained, so the time execution of TRSM is the highest.

However, it is worth noting that TRSM and AHP-TOPSIS are applicable to the QoS attribute with single value, but they are not well applicable to the QoS attribute with interval value, and they both assign weights of QoS attributes based on subjective preference. TAM can handle the above problems well.

\section{Conclusion}
In this paper, we propose an assessment and selection framework for trustworthy cloud service, which facilitate PCCs to select a trustworthy cloud service based on their actual requirements for QoS attributes of cloud service. A trustworthy cloud service selection component is designed in this framework to receive assessment requests initiated by PCCs and return assessment results to them. In order to accurately and efficiently evaluate the trust level of cloud services, a QoS based trust model is proposed and employed in this component. Such a model presents a trust level evaluation method based on the interval multiple attribute and an objective weight assignment method based on the deviation maximization to determine the trust level of cloud services provisioned by candidate CSPs. The effectiveness of the  framework is verified by a case study using a real-world dataset. The performance advantage of the trust assessment model is validated by the simulation experiment of time complexity analysis and comparison. The experimental result of a case study with an open source dataset show that the proposed trust model is effective in cloud service trust assessment and the help PCCs select a trustworthy CSP.

As future work, we aim to develop a prototype for our framework and implement it in the real environment to further verify its effectiveness.

\section*{Acknowledgements}
This work was supported by the

\ifCLASSOPTIONcaptionsoff
\newpage
\fi

\bibliographystyle{IEEEtran}
\bibliography{IEEE-FASTCloud_revision}

\end{document}